Dubininw2.doc

# Dubinin and the Brouers-Sotolongo family isotherms.

F.Brouers (a,b) and F.Marquez-Montesino (b)

a. Department of Chemical Engineering, Liege University, Belgium.

fbrouers@ulg.ac.be

b. Centro de Estudio de Energia y Tecnologias Sostenibles, University of Pinar del Rio, Cuba.

Abstract

We show how the Dubinin isotherm and its extensions can be related to the isotherms derived from the general Brouers-Sotolongo isotherm. We compare them using benzene vapor adsorbed on activated carbon data from one of the original Dubinin's paper. We use the same procedure to analyze data from the thesis of Marquez-Montesinoon the adsorption on activated carbon prepared from pinus caribaea saw dust. We conclude by proposing a simple methodology to determine the macroscopic information given by genuine statistical isotherms.

## 1.Introduction.

Adsorption isotherm is a curve giving the functional relationship between adsorbate and adsorbent in a constant-temperature adsorption process. Since the discussion is quite general and can be applied to adsorption (a physical process), chemi-sorption and bio-sorption, in the following we will use, when we do not mention explicitly adsorption, the more general term "sorption". One measures the uptake (gas volume or sorbed quantity) as a function of the gas pressure or the solution concentration). Isotherms have been used for decades in the gaseous or aqueous phase with two main objectives, first to obtain information on the nature of the sorbent surface of porous materials (sorption energy, type of porosity and type of heterogeneity) in order to use them, taking advantage of their large specific surface, and secondly to prepare and characterize specific sorbent to eliminate particular molecules in the treatment of water and more generally for the purpose of physical, chemical and biological decontamination. One usual practice is to fit the data to empirical or semi-empirical models in order to elect the best one. Then to use other empirical rules to have access to microscopic and mesoscopic information (type of porosity, sorption energy distribution) and thermodynamic quantities. This is not an easy task because, as discussed theoretically [1-4], macroscopic measurements are generated by the extreme value distribution of microscopic quantities. They can give general scaling tendencies, but cannot yield valid detailed micro or meso information without establishing correlations with independent measurements and direct micrographic analysis.

The long history of sorption is dominated by three names, Freundlich, Langmuir and Dubinin [5-7]. Freundlich [5] showed that the Henry law which predicts a linear behavior between uptake and volume or concentration for initial sorption, and which results from classical thermodynamic, was rarely obeyed and introduced in the law an exponent which he related to the heterogeneity of the surface. This is one of the first apparition, in that field, of the concept of fractality used nowadays in chemical and physico-chemical reactions [1-4, 8-13]. Langmuir [6] using the law of mass action derived his famous isotherm for surface supposed to have a unique sorption energy. Since this is a situation rarely encountered, a number of generalizations, most of them purely empirical, were introduced in the literature. In the case of ultra-porous activated carbon, Dubinin [7] introduced an empirical isotherm which has various variants and using some correlations with geometric observations and statistical hypothesis related this isotherm to the pore distribution.

Ten years ago [3], in order to give a more fundamental basis to the isotherm models and following the work of Zeldovitch [14], we show that the Freundlich exponent was related to the value and the distribution of energies contributing to the sorption, that the power law of the Freundlich isotherm could be obtained by assuming that the scale T-dependent b coefficient of the Langmuir isotherm (eq.4) was distributed as a heavy tail Lévy distribution. In that way, the sorption mechanism was viewed as a birth-death (sorption-desorption) mechanism dominated by the highest values of the random sorption energies. We showed that the exponent of the Freundlich law which appears in generalizations of the Langmuir isotherm (under the generic name "Freundlich-Langmuir isotherms"), was related to the average sorption energy and the width of its distribution. We proposed as a consequence of the discussion, to use a Weibull function as a realistic isotherm which since, has been named Brouers-Sotolongo isotherm (BS) and has been used with some success for example in porous activated carbon and carbon nanotubes [15-20], biosorption [21-22], water treatment [22-24], bacteriology [25-26], geology studies [27].

Recently extending the scope of that paper [28], we demonstrated that some of the commonly used isotherms belong to the family of a generalized Brouers-Sotolongo (GBS) obtained by replacing the exponential in the Weibull function by a deformed exponential, a function now extensively used in econometry, ecology, hydrology and many other complex systems. This function is known in the literature as BurrXII-Singh-Madalla [29-30] or q-Weibull and is a natural extension to natural and physico-chemical systems of the classical concept of exponential and appears naturally in attempts to generalize the classical thermodynamic to complex systems [31,32]. In this paper we want to demonstrate that the Dubinin isotherms belong asymptotically to the same family and analyze the various approximations using data on benzene taken from one of the original Dubinin paper[7] as well as the complete set of data of the Ph.D thesis of F.Marquez-Montesino performed in the Department of Chemical Engineering of the University of Malaga (Spain) [35,36].

## 2.The generalized Brouers-Sotolongo isotherm

In reference [28], based on statistical and mathematical considerations, we introduce the Generalized Brouers-Sotolongo (GBS) isotherm :

$$\frac{W_{GBS}}{W_{max}} = 1 - [1 + c(K/b)^{a})]^{-1/c} \quad (1)$$

Where $\kappa$ is the sorbate pressure or concentration, $W$ the up-take and $W_{max}$ , the saturation up-take in appropriate units.The coefficients c and a are form parameters and b is a scale factors. The r.h.s. of the equation is the Burr-Singh-Maddala cumulative probability distribution function. The knowledge of a, b and c allows the calculation of the usual statistical quantities of the distribution. It obeys a birth and death differential equation which is discussed in [13,33].

As shown in detail in [28], some of the most popular empirical and semi-empirical isotherms can be derived simply from the GBS isotherm, others are purely empirical such as the Redlich-Peterson isotherm. They are not correct asymptotically and, in our opinion, should be discarded.

For c= 0, one gets the Brouers-Sotolongo isotherm:

$$\frac{W_{BS}(\kappa)}{W_{max}} = 1 - \exp((\kappa/_b)^a)) \quad (2)$$

For c=1, one gets the Sips-Hill isotherm:

$$\frac{W_S(\kappa)}{W_{max}} = \frac{(\kappa/_b)^a}{1+(\kappa/_b)^a} \quad (3)$$

For c=1, a=1, one recovers the Langmuir isotherm:

$$\frac{W_L(\kappa)}{W_{max}} = \frac{(\kappa/_b)}{1+(\kappa/_b)} \quad (4)$$

These three isotherms, when $\kappa \to 0$ , give the Freundlich isotherm:

$$\frac{W_F(\kappa)}{W_{max}} = (\kappa/b)^a \quad (5)$$

**3. The Dubinin isotherm is a modified form of the Freundlich isotherm.**

According to Brouers et al. [3] the Freundlich isotherm can be understood as the asymptotic expression for low $\kappa$ (pressure or concentration) approximation of heterogeneous Langmuir isotherms (eq. 2, 3)

$$W = W_0 \kappa^{a_B} \qquad \alpha_B = \frac{\lambda\ RT}{<E_B>} \quad (6)$$

where $< E_B >$ is the average sorption energy and $\lambda$ a factor of order 1 depending on the distribution of the Langmuir isotherm scale parameter of $b$ (eq.4) see eq.25 in ref. [3]:
We can transform (6) as follows:.

$$W = W_0 \kappa^{\frac{\lambda\ RT}{<E_B>}} \quad (7)$$

$$W = W_0 \exp(ln(\kappa^{\frac{\lambda\ RT}{<E_B>}})) \quad (8)$$

$$W = W_0 \exp\left(- a_B \ln(\frac{1}{\kappa})\right) \quad (9)$$

Where $\lambda \gtrsim 1$

If $\kappa$ is a relative pressure one has:

$$W = W_0 \, exp - ( \, a_B \ln \left(\frac{p_0}{p}\right)) \quad (10)$$

The Dubinin [7] equation reads

$$W_D = W_0 \, exp - (( \, a_D \ln \left(\frac{p_0}{p}\right))^n) \qquad \text{with } \alpha_D = \frac{RT}{E_0} \quad (11)$$

Where $E_0$ is a characteristic adsorption energy and n a purely empirical factor of order 1 (1.5<n<3).

Dubinin-Radushkevich, for the case of benzene [7], have used n=2:

$$W_{DR} = W_0 \exp(-(a_D \ln \left(\frac{p_0}{p}\right))^2 \, ) \quad (12)$$

With this formula, Dubinin has established a series of relations with the microscopic geometric and energetic distribution of the micro-pores and meso-pores of the sorbent.

The comparison between equations (8) and (11) shows that the Dubinin equation is an empirical extension of the Freundlich isotherm and therefore can be related to the family of GBS isotherms (1).

## 4. **Comparison of isotherms in the case of benzene.**

In the vast literature on sorption two groups of theories have emerged. One mostly influenced by Soviet scientists works using the Dubinin formalism essentially for activated carbons, the other using Langmuir theory and its extensions. Since nowadays exist on the market efficient nonlinear regression numerical programs, a comparison of these theories, which as this has been demonstrated in the previous sections are not unrelated, is in order. We present now such comparison for original data of Dubinin on the adsorption of vapor of benzene onto an activated carbon presenting two types of porosities ultra-micro and micro porosity (Ref 7. Table 4).

In that paper, Dubinin extended its original formula to treat simultaneously the two types of porosity (D2) (micropores and ultramicropores):

$$W_{D2} = m_1 \exp\left(-\alpha_{D1} \, ln\left(\frac{p_0}{p}\right)\right)^{n_1} + m_2 \exp\left(-\alpha_{D2} \, ln\left(\frac{p_0}{p}\right)\right)^{n_2} \quad \text{with } \alpha_{Di} = RT/E_{0Di} \quad (13)$$

The values chosen to fit the experimental data were:

$$m_1 = 0.2\frac{cm^3}{g}, m_2 = 0.3\frac{cm^3}{g}, \; E_{0D1} = 0.25 \, kJ/mole, \; E_{0D2} = 0.25 \, kJ/mole,$$

corresponding to $\alpha_{D1} = 0.1$ , $\alpha_{D2} = 0.2$.

We will use the capabilities of non-linear fitting methods and we will consider the parameters of equation (13) as free parameters which will be determined by a regression method in order to compare the two families of isotherms we have used extensions of the Sips-Hill and Brouers-Sotolongo isotherms to the double porosity Dubinin's model.

The results of the numerical calculations show that in this instance as in many others due to the difficulty of determining the coefficient $c$ for sorption data, which are unique sets with a limited number of experimental points, there is practically no difference between Sips-Hill (c=1) and Brouers-Sotolongo (c=0) isotherms. In that case the recommended practice [21,34] is to use an intermediate expression the so-called Brouers-Gaspard (BG) isotherm (c=0.5)

$$\frac{W_{BG}}{W_{max}} = 1 - [1 + 2(\kappa/b)^{a} )]^{-2} \quad (14)$$

which we have extended to the two-porosity model (BG2).

$$W_{BG2} = m_1 \left(1 - [1 + 2(\kappa/b_1)^{a_1} )]^{-2} \right) + m_2(1 - [1 + 2(\kappa/b_2)^{a_2} )]^{-2} ) \quad (15)$$

with $\kappa = \frac{p}{p_0}$ and $\alpha_i = \frac{\lambda \ RT}{<E_i>}$

Here are the results:

In Fig.1 we present the fitting of the benzene data of (ref. 7, table 4) with simple Dubinin (eq. 11) isotherm (black), Langmuir (light black), Brouers Sotolongo (dashed) , Sips-Hill isotherms (dashed).

In Table 1a we show the results for the one porosity Dubinin and BG. isotherm (intermediate between BS and Sips. We can see that the BG isotherm is more precise that the simple Dubinin isotherm (D1). It appear that the one-pore Dubinin equation does not fit the data while the Sips-Hill, Brouers-Sotolongo and the intermediate Brouers-Gaspard practically coincide with the data curve.

In Fig.2 we present the double porosity Dubinin isotherm (D2) eq.13 (dotted) and double porosity BG2 (15) isotherm (dashed).They cannot be distinguished and fit both perfectly the data.

In Table 1b we report the results for the two-porosity model. As this can be observed, there is no significant difference between the two approaches and the Dubinin original assumed values of the coefficients. For the two-pores case the values of the parameters chosen by Dubinin are recovered with very small change and great accuracy with the Dubinin two pores equation and the intermediate Brouers-Gaspard isotherm using the “mathematica” recursion program.

The energies obtained are different: in the Dubinin formalism, it is the characteristic adsorption energy. In the BG formalism the energy is related to the averaged sorption energy depending on the energy distribution.

The fact that $n_1$ and $n_2$ are larger than $n$ is consistent with the observation [35] that in the case of activated carbons $n$ varies from 3 to 1 as the system becomes more heterogeneous.

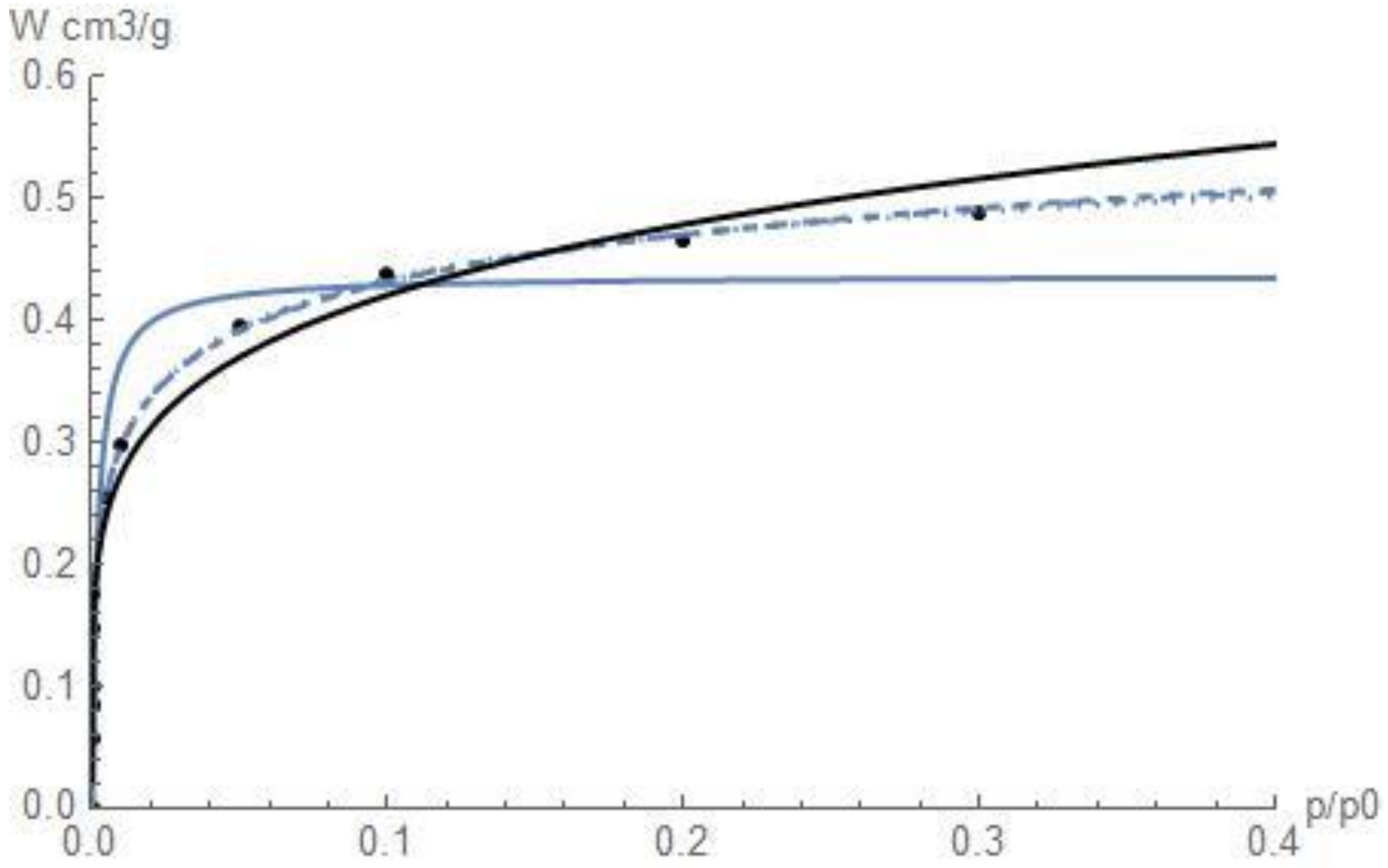

In Fig.1 we present the fitting of the benzene data of (ref.7 table 4) with simple one porosity Dubinin D1 (eq. 12 ) isotherm (black) and one porosity Langmuir (light black), Brouers Sotolongo (dashed) and Sips-Hill isotherms (dot-dashed).

| | $m$ | $a_D$ | $a_B$ | $E_{0D}$ kJ/mol | $< E >$ kJ/mol | $n$ | $R^2$ |
|---|---|---|---|---|---|---|---|
| D1 | 0.646 | 0.131 | | 1.86 | | 1.42 | 0.996 |
| BG | 0.654 | | 0.292 | | 7.92 $\lambda$ | | 0.999 |

Table 1a Results of the one porosity Dubinin D1 (eq.12) and one porosity Brouers-Gaspard isotherms (eq.14) for the benzene data of Ref.7 Table 4

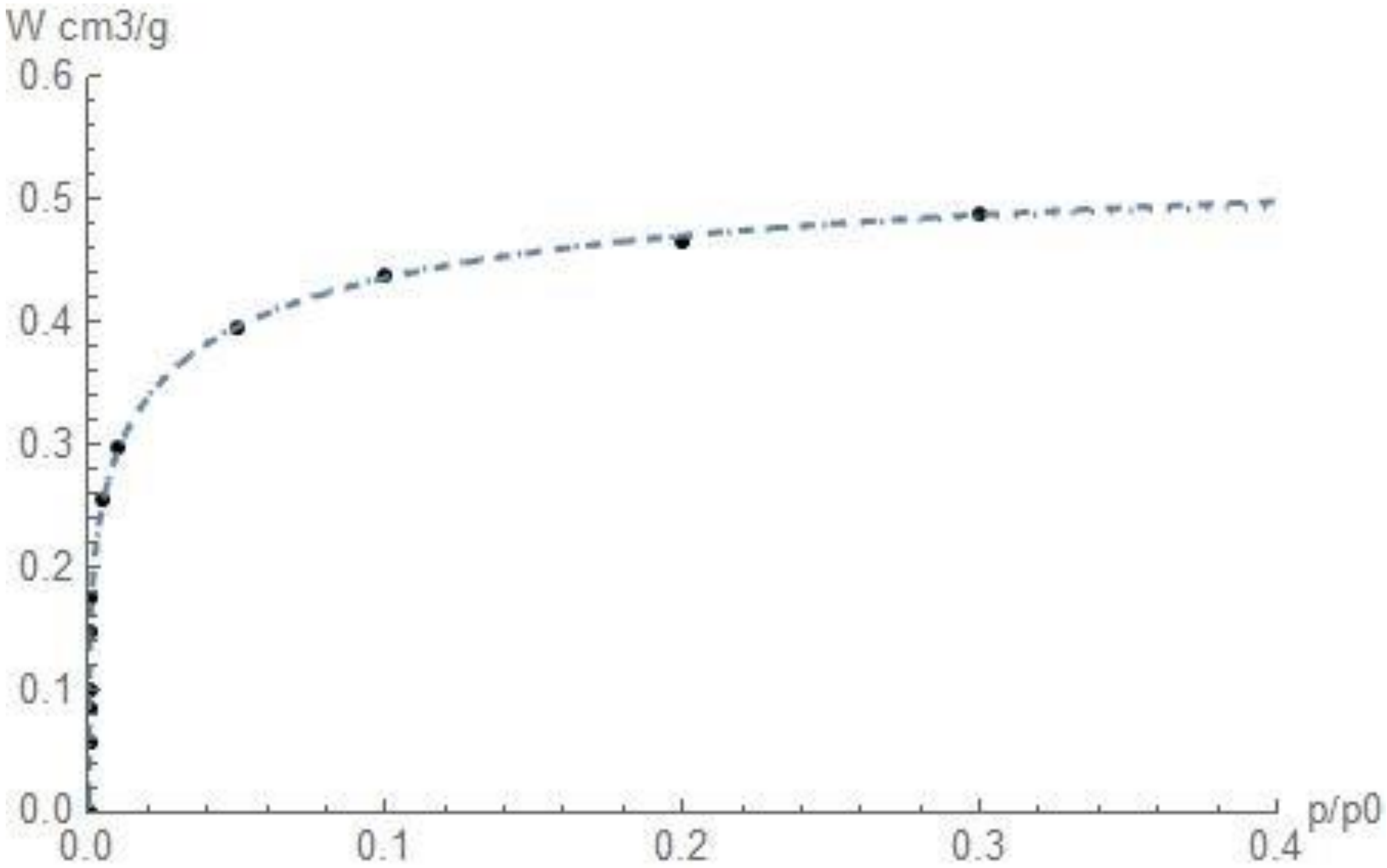

In Fig.2 we present the double porosity Dubinin D2 isotherm eq.13 (dotted) and double porosity BG2 isotherm eq. 15 (dashed).They cannot be distinguished and fit perfectly the data.

| | $m_1$ | $m_2$ | $E_{D1}$ kJ/mol | $E_{D2}$ kJ/mol | $< E_1 >$ kJ/mol | $< E_2 >$ kJ/mol | $n_1$ | $n_2$ | $R^2$ |
|---|---|---|---|---|---|---|---|---|---|
| D2 | 0.289 | 0.218 | 24 | 12 | | | 2.02 | 1.95 | 0.99996 |
| BG2 | 0.322 | 0.239 | | | 8.75 $\lambda$ | 4.90 $\lambda$ | | | 0.99994 |
| D1980 | 0.3 | 0.2 | 25 | 12.5 | | | 2.00 | 2.00 | |

Table 1b results of the two porosity Dubinin isotherm (D2, eq.13), the two porous Brouers-Gaspard isotherm (BG2,eq.15 ) and the original Dubinin results (Ref.7 Table 4)

**5. Analysis of the Marquez thesis data.**

We will now analyze with isotherms (2-5 and 11) and discuss the complete set of data from one of us [35,36] on powered activated carbon from "pinus caribea" saw dust with the use of $CO_2$ and water vapor as activation agent

Here again, it is difficult to make a distinction between Brouers-Sotolongo and Sips-Hill isotherms and to construct the table and observe the correlation between isotherms coefficients and physical characteristics we have use the intermediate (c=0.5) BG isotherm (eq.14). The results obtained with that isotherm are compared with the physical and isotherm data reported in the work of Marquez who had used the Dubinin-Radushkevich (n=2) formula.

The results reported here show that there is practically no difference between the two methods. The only one arises from the different interpretation of the obtained energy: the adsorption energy in the Dubinin's approach and a estimation of the average adsorption energy in the second approach. The results are summarized in the Fig3 and the Tables 2- 4 In order to find some useful correlations, the isotherms parameters are presented with the corresponding physical data (degree of activation, BET surface, porosity ) taken directly from ref [36].

In order to illustrate these conclusions in Fig.3 we show the fitting of the case with the smallest $R^2$ (0.99997).

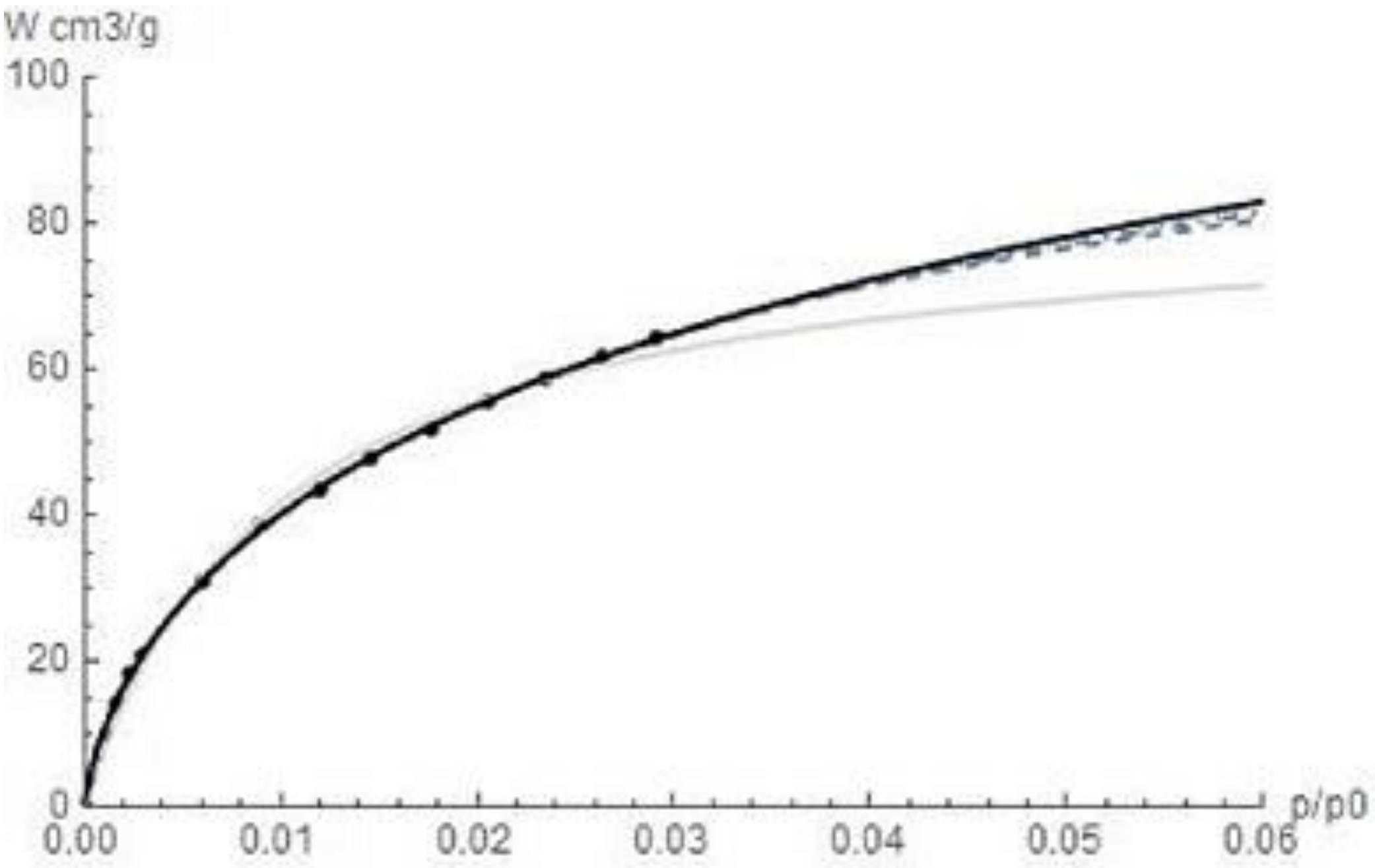

In Fig.3 Fit with the Dubinin isotherm (black ), the Brouers-Gaspard isotherm (dashed) and the Freundlich isotherm (grey) corresponding CA S72547 ($R^2$ =0.99997). They practically coincide with the data. All other fittings have a higher $R^2$

The fitting with the recursion "mathematica" program gives the following complete set of results? The values are taken from ref. [35,36].

| C.A./ $H_20$ | $a$ | $b$ | $W_m$ | $R^2$ | $s.BET \quad m^2/g$ | $burn\ off\%$ |
|---|---|---|---|---|---|---|
| S72527 | 0.66 | 0.022 | 99.37 | 0.99998 | 570 | 27 |
| S72547 | 0.67 | 0.034 | 123.01 | 0.99997 | 732 | 41 |
| S72561 | 0.67 | 0.045 | 139.57 | 0.99999 | 895 | 61 |
| S72571 | 0.66 | 0.239 | 268.25 | 0.99998 | 1038 | 71 |

| C.A./ $H_20$ | $E_{0D}$ kJ/mol | $< E_{BG} >$ kJ/mol | W(0.03) $cm^3/g$ | $micro\ p.$ $cm^3/g$ | $meso\ p.$ $cm^3/g$ | $Total\ p.$ $cm^3/g$ |
|---|---|---|---|---|---|---|
| S72527 | 12.64 | 3.43 $\lambda$ | 62. | 0.28 | 0.15 | 1.05 |
| S72547 | 11.70 | 3.38 $\lambda$ | 64.9 | 0.31 | 0.23 | 1.15 |
| S72561 | 11.47 | 3.38 $\lambda$ | 65.8 | 0.37 | 0.35 | 1.68 |
| S72571 | 11.14 | 3.73 $\lambda$ | 57.9 | 0.43 | 0.52 | 2.76 |

| C.A./ $H_20$ | $a$ | $b$ | $W_m$ $cm^3/g$ | $R^2$ | $s.BET$ $m^2/g$ | $burn\ off$ %. |
|---|---|---|---|---|---|---|
| S85029 | 0.66 | 0.027 | 107.51 | 0.99997 | 617 | 29 |
| S85049 | 0.67 | 0.042 | 157.54 | 0.99998 | 790 | 49 |
| S85068 | 0.68 | 0.094 | 190.42 | 0.99999 | 937 | 60 |
| S85075 | 0.70 | 0.0.98 | 186.10 | 0.99997 | 1034 | 76 |

| C.A./ $H_20$ | $E_{0D}$ kJ/mol | $< E_{BG} >$ kJ/mol | W ( 0.03) $cm^3/g$ | $micro\ p.$ $cm^3/g$ | $meso\ p.$ $cm^3/g$ | $Total\ p.$ $cm^3/g$ |
|---|---|---|---|---|---|---|
| S85029 | 12.58 | 3.43 $\lambda$ | 61.5 | 0.27 | 0.11 | 0.97 |
| S85049 | 1156 | 3.38 $\lambda$ | 67.3 | 0.34 | 0.30 | 1.46 |
| S85068 | 10.85 | 3.38 $\lambda$ | 64.2 | 0.38 | 0.58 | 1.65 |
| S85075 | 10.73 | 3.84 $\lambda$ | 57.4 | 0.39 | 0.60 | 3.02 |

Tabla 2. Results of Brouers-Gaspard and Dubinin isotherms analysis of carbon activated with water vapor (data from ref. [35,36] and parameters from eq. 11 and 14)

| C.A./ $CO_2$ | $a$ | $b$ | $W_m$ $cm^3/g$ | $R^2$ | $ss.BET$ $m^2/g$ | $burn\ off$ % |
|---|---|---|---|---|---|---|
| C75029 | 0.58 | 0.049 | 161.80 | 0.99999 | 575 | 29 |
| C75061 | 0.70 | 0.054 | 207.34 | 0.99999 | 969 | 61 |
| C75077 | 0.71 | 0.082 | 245.93 | 0.99999 | 1021 | 77 |

| C.A./ $CO_2$ | $E_{0D}$ kJ/mol | $< E_{BG} >$ kJ/mol | W(0.03) $cm^3/g$ | $micro\ p.$ $cm^3/g$ | $meso\ p.$ $cm^3/g$ | $Total\ p.$ $cm^3/g$ |
|---|---|---|---|---|---|---|
| C75029 | 14.47 | 3.91 $\lambda$ | 76.5 | 0.27 | 0.064 | 0.76 |
| C75061 | 12.17 | 2.24 $\lambda$ | 90.8 | 0.46 | 0.10 | 1.26 |
| C75077 | 11.12 | 3.18 $\lambda$ | 87.4 | 0.48 | 0.17 | 1.15 |

| C.A./ $CO_2$ | $a$ | $b$ | $W_m$ $cm^3/g$ | $R^2$ | $s.BET$ $m^2/g$ | $burn\ off$ %. |
|---|---|---|---|---|---|---|
| C87528 | 0.65 | 0.020 | 112.78 | 0.99999 | 567 | 28 |
| C87548 | 0.67 | 0.028 | 145.50 | 0.99998 | 739 | 48 |

| C.A./ $CO_2$ | $E_{0D}$ | $< E_{BG} >$ kJ/mol | W (0.03) $cm^3/g$ | $micro\ p.$ $cm^3/g$ | $meso\ p.$ $cm^3/g$ | $Total\ p.$ $cm^3/g$ |
|---|---|---|---|---|---|---|
| C87528 | 19.00 | 3.49 $\lambda$ | 71.6 | 0.36 | 0.075 | 0.84 |
| C87548 | 13.8 | 3.38 $\lambda$ | 83.3 | 0.38 | 0.23 | 1.38 |

Tabla 3. Results of Brouers-Gaspard and Dubinin isotherms analysis of carbon activated with $CO_2$ (data from ref. [35,36] and parameters from eq. 11 and 14)

## 6. Analysis of the results of Tables 2 and 3.

One cannot observe significant variation of the fractal parameter a. This would indicate that the distribution energy width is practically the same in all cases since the temperature is constant during adsorption measurements. The two other parameters, $W_{max}$and the scale $b$ increases regularly with both % of burn-off and BET specific surface but the adsorption at the maximum relative pressure W(0.03) does not show much variation. The Dubinin characteristic adsorption energy $E_0$ does not change much either. The same for the average energy <E> varying slightly in accordance with $E_0$.

## 7. Conclusions.

The results of our paper show that the Brouers-Sotolongo and the Sips-Hill isotherms which are among the many semi-empirical formulas used in the literature and which are the only one having the good asymptotic behavior and genuine statistical distribution properties [28] give almost undistinguishable results for this type of systems. That important conclusion has been observed in many instances where the isotherm is measured until saturation. The suggestion has been made in practice to use an intermediate isotherm (Brouers-Gaspard isotherm introduced in [21]) by giving the intermediate value ½ to the complexity parameter c in eq.1. It is with this isotherm that we have treated data of ref. [7] from Dubinin and those of [35,36] measured in the Department of Chemical Engineering of the University of Malaga (Spain). We get very close results if we use both Dubinin and Brouers-Sotolongo formalisms. The macroscopic parameters derived from these isotherms are complementary and are compared with the physical characterization of the carbon.

To conclude we suggest that in usual practice it would be sufficient to restrict oneself with the following isotherms Dubinin (eq.11) Sips-Hill and Brouers-Sotolongo (eq.14). If in one series of data it is not possible to choose between these two last isotherms, for the sake of comparison, as this is done in this paper, one can choose the intermediate Brouers-Gaspard isotherm (eq.14). Non-linear regression methods are a must. The comparison with a great number of empirical isotherms and the use of imprecise linearization methods has become obsolete. We must remember that the isotherms can give only macroscopic information and the correlation of the isotherms parameters with the micro- or meso- pore geometry requires independent methods like micrography and other physical methods. Only in that way empirical correlations between mico-structure and sorption measures can be done. It must be remembered that the statistics of isotherms are not very precise. Generally one makes a unique measurement for every choice of physical parameters (T, pH, concentration, …), there is no information on the error-bars, the measurements depend on the experimental conditions [37] and are the result of multi-scale averages.

Acknowledgments.

F.B. is grateful to Professor Jean-Paul Pirard (Liege) who introduced him to the world of adsorption, to Professors Sarra Gaspard (Guadeloupe, France), Mongi Seffen (Sousse, Tunisie) and Al-Musawi (Bagdad, Irak) for useful discussions.